\documentclass{emulateapj}

\usepackage{natbib}
\usepackage{amsmath}
\usepackage{color}

\shorttitle{Progenitors of Recombining SNRs}
\shortauthors{Moriya}

\begin{document}

\title{
PROGENITORS OF RECOMBINING SUPERNOVA REMNANTS
}

\def\ipmu{1}
\def\ut{2}
\def\resceu{3}

\author{
{Takashi J. Moriya}\altaffilmark{\ipmu,\ut,\resceu}
}

\altaffiltext{\ipmu}{
Kavli Institute for the Physics and Mathematics of the Universe,
Todai Institutes for Advanced Study,
University of Tokyo, Kashiwanoha 5-1-5, Kashiwa, Chiba 277-8583, Japan;
takashi.moriya@ipmu.jp
}
\altaffiltext{\ut}{
Department of Astronomy, Graduate School of Science, University of Tokyo,
Hongo 7-3-1, Bunkyo, Tokyo 113-0033, Japan
}
\altaffiltext{\resceu}{
Research Center for the Early Universe,
Graduate School of Science, University of Tokyo,
Hongo 7-3-1, Bunkyo, Tokyo 113-0033, Japan
}

\begin{abstract}
Usual supernova remnants have either ionizing plasma or plasma in collisional
ionization equilibrium, i.e., the ionization temperature
is lower than or equal to the electron temperature.
However, the existence of recombining supernova remnants, i.e.,
supernova remnants with the ionization temperature higher than
the electron temperature, is recently confirmed.
One suggested way to have recombining plasma in a supernova remnant is to have
a dense circumstellar medium at the time of the supernova explosion.
If the circumstellar medium is dense
enough, collisional ionization equilibrium can be established in the
early stage of the evolution of the supernova remnant and subsequent
adiabatic cooling which occurs after
the shock wave gets out of the dense circumstellar
medium makes the electron temperature lower than the ionization temperature.
We study the circumstellar medium around several supernova progenitors
and show which supernova progenitors can have a circumstellar medium
which is dense enough to establish collisional ionization equilibrium
soon after the explosion.
We find that the circumstellar medium around red supergiants 
(especially massive ones) and
the circumstellar medium which is dense enough to
make Type IIn supernovae can establish collisional ionization equilibrium
soon after the explosion and can evolve to recombining supernova remnants.
Wolf-Rayet stars and white dwarfs have possibility to be
recombining supernova remnants but the fraction is expected to be very small.
As the occurrence rate of the explosions of red supergiants is much higher
than that of Type IIn supernovae, the major progenitors of recombining supernova
remnants are likely to be red supergiants.
\end{abstract}


\section{Introduction}
X-ray observations by the {\it Suzaku} satellite are
confirming the existence of recombining supernova remnants (SNRs)
\citep{yamaguchi2009,yamaguchi2012,ozawa2009,ohnishi2011,sawada2012}.
Recombining SNRs are SNRs in which the ionization temperature is higher
than the electron temperature.
The forward shock wave emerged at the time of a supernova (SN) explosion
propagates in the interstellar medium (ISM).
As the typical density of the ISM is very small
($n_e\sim 1~\mathrm{cm^{-3}}$ or less where $n_e$ is the electron number density),
the timescale to reach collisional ionization equilibrium (CIE) in
the shocked ISM is typically $\sim 10^4$ years or longer
\citep[e.g.,][]{masai1984}.
Electrons heated by the Coulomb interaction with
ions in the shocked ISM collisionally excite ions
and reach CIE with this timescale.
Thus, young SNRs before CIE are supposed to be ionizing SNRs in which the electron
temperature is higher than the ionization temperature and evolve
to SNRs in CIE. Most of SNRs are known to be in either the ionizing
stage or CIE \citep[e.g.,][]{kawasaki2005}. 
In this simple picture, SNRs cannot be recombining SNRs and the
confirmation of the recombining SNRs challenges the current
understanding of the evolution of SNRs.

There are several suggested mechanisms to make recombining plasma
in SNRs \citep[see, e.g.,][and references therein]{yamaguchi2012}.
The existence of a dense circumstellar medium (CSM) is
one possible way to explain the recombining SNRs
\citep[e.g.,][]{itoh1989,shimizu2011,zhou2011}.
If a dense CSM is around a SN,
CIE can be achieved in much shorter timescale
$[\sim 10^{4}/(n_e/1~\mathrm{cm^{-3}})~\mathrm{years}]$.
When the shock wave reaches the outer edge of the dense CSM,
the shocked CSM suddenly expands adiabatically and the
electron temperature suddenly becomes low and
the plasma starts to recombine.

Although the existence of the dense CSM at the time of the SN explosion
has been suggested as a possible mechanism to realize recombining SNRs,
we still do not have a clear picture about possible SN progenitors that
can have a CSM which is dense enough to make recombining SNRs.
Here, in this {\it Letter}, we look into the properties of
the CSM around SN progenitors at the time of SN explosions
and investigate the progenitors which can evolve
to recombining SNRs.
We focus on massive star progenitors because two recombining
SNRs, IC 443 and W49B, are clearly associated with massive star forming
regions \citep[e.g.,][]{yamaguchi2012} but we also investigate
possible channels for white dwarfs to be recombining SNRs.

\section{Possible Progenitors}
\subsection{Red Supergiants \& Wolf-Rayet Stars}
Red supergiants (RSGs) and Wolf-Rayet (WR) stars are progenitors of core-collapse SNe.
Because of their high luminosities, they lose their mass
before their explosions. Thus, RSGs and WR stars explode
inside the CSM created by the preceding stellar evolution.
If we assume that the CSM is from a steady wind with 
the velocity $v_w$ and the mass-loss rate $\dot{M}$,
the wind density $\rho_w$ becomes
\begin{equation}
\rho_w =\frac{\dot{M}}{4\pi r^2 v_w}, \label{CSMrho}
\end{equation}
where $r$ is the radius.

If the star inside the CSM explodes, a forward shock propagates in
the CSM.
Assuming that the adiabatic index of the system is $3/5$,
the density $\rho_s$ of the shocked CSM just behind the forward shock becomes
$\rho_s=4\rho_w$. 
As the forward shock propagates in the CSM, it is decelerated,
especially if the CSM is dense. However, 
the mass of the CSM swept up by the forward shock is still small
compared to the progenitor mass in the early epochs we are interested in
and we assume that it is freely expanding with the velocity $v_s$ for
simplicity. Note that the deceleration makes the time of the
interaction between the shock wave and the CSM longer and
CIE can be achieved easier with the deceleration.
The typical $v_s$ of standard SN explosions is
$v_s\sim 10,000~\mathrm{km~s^{-1}}$ \citep[e.g.,][]{suzuki1995,fransson1996,dwarkadas2005,dwarkadas2007}.
The location of the forward shock at the time $t$ after the
explosion is $r=v_st$ and $\rho_s$ can be expressed as
\begin{equation}
\rho_s=\frac{\dot{M}}{\pi v_s^2t^2v_w}. \label{shockrho}
\end{equation}
Although Equation (\ref{shockrho}) is the evolution of the 
density just behind the shock, the remaining entire shocked CSM has
similar densities when the shock is traveling
in the density structure close to $\rho_w\propto r^{-2}$
\citep[see, e.g.,][]{chevalier1982,suzuki1995,fransson1996,dwarkadas2005,dwarkadas2007}
and we assume that $\rho_{s}$ is a typical value in the shocked CSM.
The actual densities in the shocked CSM are slightly
higher than $\rho_{s}$.

Since the wind properties of RSGs and WR stars differ,
we consider two cases separately.

\begin{figure}
\begin{center}
\includegraphics[width=\columnwidth]{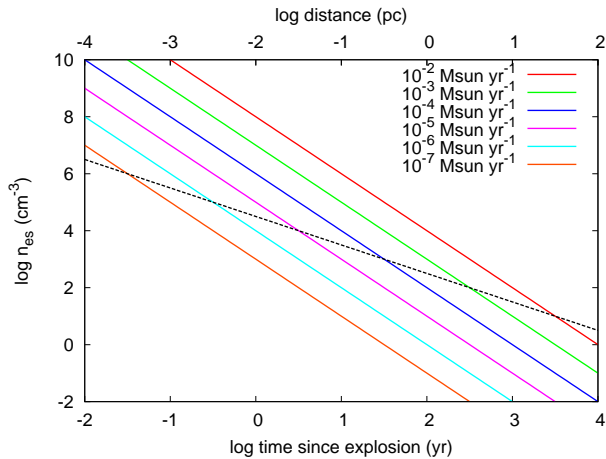}
\caption{
Evolution of the electron number density in the shocked CSM from
RSG winds.
$v_s=10,000~\mathrm{km~s^{-1}}$ and $v_w=10~\mathrm{km~s^{-1}}$
is fixed and lines with several mass-loss rates are shown.
The distance in the horizontal axis is the distance
which can be reached by the forward shock with
$v_s=10,000~\mathrm{km~s^{-1}}$ in the corresponding time.
If we assume that the progenitor is in the RSG stage for $10^5$ years
before the explosion, the CSM can reach $1$ pc with $v_w=10~\mathrm{km~s^{-1}}$.
The dashed line follows $n_{es}t=10^{12}~\mathrm{cm^{-3}~sec}$ and
corresponds to the minimum density required to reach CIE at the time.
Shocked CSM above this line is presumed to be at CIE.
}\label{rsg}
\end{center}
\end{figure}

\subsubsection{Red Supergiants}
The typical mass-loss rate and wind velocity of RSGs are
$\sim 10^{-5}~M_\odot~\mathrm{yr^{-1}}$ and
$\sim 10~\mathrm{km~s^{-1}}$, respectively \citep[e.g.,][]{mauron2011}
and they are consistent with those estimated from the observations
of SN explosions from RSGs \citep[Type IIP SNe, e.g.,][]{chevalier2006}.
If we assume that the RSG wind has the solar metallicity and H and He
in the wind are fully ionized when the forward shock passes,
$\rho_s=2.0\times 10^{-24}n_{es}$ where $n_{es}$ is the electron number
density in the shocked CSM. From Equation (\ref{shockrho}),
the time evolution of $n_{es}$ is
\begin{equation}
n_{es}t^2= 10^{20}\dot{M}_{-5}v_{s,9}^{-2}v_{w,6}^{-1}~\mathrm{cm^{-3}~sec^{2}},
\label{rsgnes}
\end{equation}
where $\dot{M}_{-5}$ is $\dot{M}$ scaled by $10^{-5}~M_\odot~\mathrm{yr^{-1}}$,
$v_{s,9}$ is $v_s$ scaled by $10,000~\mathrm{km~s^{-1}}$, and
$v_{w,6}$ is $v_w$ scaled by $10~\mathrm{km~s^{-1}}$.

Electrons and ions in plasma can reach CIE with the timescale of
\begin{equation}
n_{es}t\sim 10^{12}~\mathrm{cm^{-3}~sec},\label{cie}
\end{equation}
\citep[e.g.,][]{masai1984,rsmith2010}.
Note that only ions are presumed to be heated by the forward shock
and electrons are heated up by the subsequent Coulomb interaction
between ions and electrons.
The timescale of the electron heating is \citep[e.g.,][]{masai1994}
\begin{equation}
n_{es}t=3\times 10^{14} v_{s,9}^3\left(\frac{\mu}{0.5}\right)^{1.5}
\left(\frac{\ln \Lambda}{30}\right)^{-1}~\mathrm{cm^{-3}~sec},
\end{equation}
where $\mu$ is the mean molecular weight and $\ln\Lambda$ is the Coulomb logarithm.
Although the timescale of temperature equilibrium is a few
orders of magnitudes longer than that of CIE,
the electron temperature can reach about 10 \% of
the ion temperature ($\sim 10^{9}$ K) in the CIE timescale \citep{masai1994}
and becomes high enough to explain the ionization temperature of recombining SNRs.
Recombining plasma in SNRs can appear if electrons cool down
after CIE is achieved \citep[e.g.,][]{itoh1989}.

Figure \ref{rsg} shows the comparison of
the evolution of the typical density in the shocked CSM (Equation (\ref{rsgnes}))
and the CIE timescale (Equation (\ref{cie})).
CIE can be achieved at early epochs of SNRs from RSGs 
with the typical mass-loss rate
if we take into account the existence of the CSM.
This is contrary to the general belief that it takes much time to be CIE
because SNRs evolve in ISM.
Note that $v_s$ in the early time is presumed to be
higher than the value assumed in Figure \ref{rsg} \citep[e.g.,][]{dwarkadas2005}
and this effect can make the evolution of the electron number density faster.
In addition, the mass of the recombining plasma
estimated from Figure \ref{rsg} in the case of the
standard mass loss is $\simeq 3\times 10^{-2}~M_\odot$ and rather small.
Massive RSGs, yellow supergiants, or
RSGs in binary systems can have higher
mass-loss rates than less massive RSGs especially just before their explosions
\citep[see, e.g.,][and references therein]{georgy2012} and they are
more likely to become recombining SNRs among RSGs.

The early X-ray observations of Type IIb SN 1993J
whose progenitor is an RSG in a binary system \citep[e.g.,][]{maund2004}
revealed the existence of the CIE plasma
at a few days since the explosion \citep[e.g.,][]{uno2002}
and the progenitor's mass-loss rate is suggested to be
$\simeq 5\times 10^{-5}~M_\odot~\mathrm{yr^{-1}}$
\citep[e.g.,][]{suzuki1995,fransson1996}.
This is consistent with our estimate and
explosions of RSGs can establish CIE in the early epochs and
evolve to recombining SNRs.

\begin{figure}
\begin{center}
\includegraphics[width=\columnwidth]{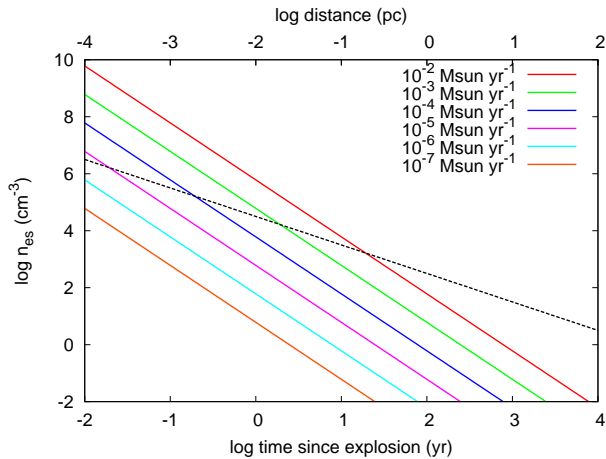}
\caption{
Evolution of the electron number density in the shocked CSM from
WR star winds.
$v_s=10,000~\mathrm{km~s^{-1}}$ and $v_w=1000~\mathrm{km~s^{-1}}$
is fixed and lines with several mass-loss rates are shown.
The distance in the horizontal axis is the distance
which can be reached by the forward shock with
$v_s=10,000~\mathrm{km~s^{-1}}$ in the corresponding time.
If we assume that the progenitor is in the WR stage for $10^5$ years
before the explosion, the CSM can reach $100$ pc with $v_w=1000~\mathrm{km~s^{-1}}$.
The dashed line follows $n_{es}t=10^{12}~\mathrm{cm^{-3}~sec}$ and
corresponds to the minimum density required to reach CIE at the time.
Shocked CSM above this line is presumed to be at CIE.
}\label{wr}
\end{center}
\end{figure}

\subsubsection{Wolf-Rayet Stars}
The typical mass-loss rate and wind velocity of WR stars are
$\sim 10^{-5}~M_\odot~\mathrm{yr^{-1}}$ and
$\sim 1000~\mathrm{km~s^{-1}}$, respectively \citep[e.g.,][]{crowther2007}.
If we assume that the wind from WR stars is composed of 50 \% of carbon and 50
\% of oxygen and they are fully ionized when the CSM is shocked,
$\rho_s=3.3\times 10^{-24}n_{es}$.
With these values, the following equation is obtained from Equation (\ref{shockrho}):
\begin{equation}
n_{es}t^2=6\times 10^{17}\dot{M}_{-5}v_{s,9}^{-2}v_{w,8}^{-1}~\mathrm{cm^{-3}~sec^{2}},
\end{equation}
where $v_{w,8}$ is $v_w$ scaled by $1000~\mathrm{km~s^{-1}}$.

The evolution of $n_{es}$ is compared to the CIE timescale
$n_{es}t\sim10^{12}~\mathrm{cm^{-3}~sec}$ in Figure \ref{wr}.
With the canonical mass-loss rate $10^{-5}~M_\odot~\mathrm{yr^{-1}}$,
the evolution of the shocked CSM can be comparable to the CIE timescale
just after the explosion.
However, $v_s$ is presumed to be larger than $v_s=10,000~\mathrm{km~s^{-1}}$
in the early epochs and the evolution of $n_{es}$ can be faster, as noted in the previous section.
Contrary to the case of RSGs, CIE can be only established at the very
early epochs with the optimistic $v_s$ and
it is likely that CIE is not achieved
at early phases of the typical explosions of WR stars with
the canonical mass-loss history.
This is because the typical wind
velocity is about 100 times larger than the typical RSG wind velocity
and the CSM becomes thin much faster.
Although recent radio observations of explosions of
WR stars (Type Ibc SNe) are revealing the existence of WR stars with
high mass-loss rates \citep[$\sim 10^{-4}~M_\odot~\mathrm{yr^{-1}}$, e.g.,][]{wellons2012},
the amount of recombining plasma is very small ($\sim 10^{-4}~M_\odot$, Figure \ref{wr})
even if such a high mass-loss rate is maintained for the entire WR phase.
Thus, WR stars may be difficult to have dense CSM which is massive
enough to make them recombining SNRs.
Since some elements can reach the CIE in smaller timescales than
$n_{es}t\sim10^{12}~\mathrm{cm^{-3}~sec}$ at the typical
temperature in the shocked CSM \citep{rsmith2010},
at least some elements may reach CIE.
In addition, it is also known that some WR stars
experience explosive mass loss just before their explosions which
can eject massive CSM, as is indicated by the progenitor of Type Ib SN 2006jc \citep[e.g.,][]{pastorello2007}.
Explosions of this kind of WR stars can also result in recombining SNRs
but they are also expected to be rare.

\subsection{Type IIn Supernova Progenitors}\label{sec:iin}
Type IIn SNe are SNe which show narrow spectral lines, especially of hydrogen,
in optical spectra \citep[e.g.,][]{schlegel1990}.
Their spectral features can be explained by the existence of the dense
CSM with the Thomson optical depth $\tau_T$ larger than $1$
\citep[e.g.,][]{chugai2001,dessart2009}.
They can become luminous in X-ray and radio and some of them become
luminous even in optical. The high luminosities of Type IIn SNe can be
naturally explained by the interaction between SN ejecta and its dense CSM
\citep[e.g.,][]{dwarkadas2010,chugai2004}
and the progenitors of Type IIn SNe are a plausible
candidate of the origin of recombining SNRs.

We assume that Type IIn SNe have a dense CSM with
$\tau_T=\sigma_T\overline{n}_{ew}\Delta R\sim 1$, where $\sigma_T$ is the
Thomson cross section, $\overline{n}_{ew}$ is the mean CSM electron
number density, and $\Delta R$ is the CSM length.
We use the mean density $\overline{n}_{ew}$ because
the mass loss of Type IIn SN progenitors just before their explosions
are revealed to be non-steady from X-ray observations \citep{dwarkadas2012}.
The shock wave with the velocity $v_s$ can propagate through the CSM with $t_s=\Delta R/v_s$.
As CSM or ISM with much lower density exists outside the dense CSM
and the shocked CSM is rarefied after the shock wave goes out of
the dense CSM, recombining plasma can be easily synthesized once the CIE is
achieved in the shocked CSM.
Assuming $\overline{n}_{es}=4\overline{n}_{ew}$ and 
the solar metallicity, the typical timescale before the rarefaction is
\begin{equation}
\overline{n}_{es}t_s\sim 6\times 10^{15} v_{s,9}^{-1}~\mathrm{cm^{-3}~sec}.
\label{iin}
\end{equation}
This is much larger than the timescale required to achieve
the CIE, $n_et\sim 10^{12}~\mathrm{cm^{-3}~sec}$.
In reality, $v_s$ can be smaller than $10,000~\mathrm{km~s^{-1}}$
because of the deceleration by the dense CSM
but this makes $t_s$ longer.
One caveat is that we use a constant mean electron density
$\overline{n}_{ew}$ to estimate the density evolution.
If the density declines very steeply, this assumption
can be very crude and Type IIn SNe from
very steep CSM may not end up with recombining SNRs.
Nonetheless, many Type IIn SNe have flat density CSM \citep{dwarkadas2012}
and we presume that most of Type IIn SNe can end up with recombining SNRs.
However, as explosions of RSGs (Type II SNe) occurs much more frequently
than Type IIn SNe \citep[e.g.,][]{li2011},
the major progenitors of recombining SNRs
are likely to be RSGs.

Unfortunately, the progenitors of Type IIn SNe are not understood well.
\citet{gal-yam2009} have confirmed that 
the progenitor of a Type IIn SN, SN 2005gl, is a luminous blue variable
\citep[LBV, e.g.,][]{humphreys1994}. As LBVs originate from very massive stars (more than
$M_\mathrm{ZAMS}\sim40M_\odot$ where $M_\mathrm{ZAMS}$
is the zero-age main-sequence mass, e.g., \citet[][]{crowther2007})
and the observational rate of Type IIn SNe is also consistent
with the mass range of LBVs \citep[e.g.,][]{nsmith2011},
Type IIn SNe are suggested to come mainly from these very massive stars.
However, it is theoretically considered that LBVs are in an evolutionary stage
in which very massive stars evolve to WR stars and LBVs do not explode.
It is also possible that the fast wind from a WR star collides to the
slowly moving wind from its previous RSG
stage and a dense shell which is enough to be a Type IIn SN
is created by the interaction
\citep[e.g.,][]{dwarkadas2010}.
Another possible Type IIn SN progenitor 
is a super-asymptotic giant branch (AGB) star with
$M_\mathrm{ZAMS}\sim 8~M_\odot$. An O+Ne+Mg core at the center of
the super-AGB wind can be an electron-capture SN \citep[][]{nomoto1984}.
The progenitor of Type IIn SN 2008S is found to be
around $10~M_\odot$ \citep{prieto2008} and may belong to this class
\citep[e.g.,][]{botticella2009}
but there also exists an argument that SN 2008S may not be an SN \citep{nsmith2009}.
Finally, we note that some of Type Ic superluminous SNe
recently discovered \citep[e.g.,][]{quimby2011}
are related to the interaction of a dense C+O-rich
CSM and SN ejecta \citep[][]{blinnikov2010,moriya2012}
and they can also be a progenitor of recombining SNRs.
However, Type Ic superluminous SNe preferentially appear in metal-poor galaxies
\citep{quimby2011,neill2011} and the occurrence rate is also quite small.
Thus, recombining SNRs currently observed in our Galaxy seem irrelevant
to them.

\subsection{White Dwarfs}
Although most of the recombining SNRs currently discovered are likely to originate
from core-collapse SNe \citep[e.g.,][]{yamaguchi2012}, Type Ia SNe can also evolve to
recombining SNRs although it is expected to be quite rare.
Type Ia SNe are explosions of white dwarfs.
There are two major suggested paths for white dwarfs
to explode:
single degenerate (SD) channel \citep[e.g.,][]{nomoto1982}
and double degenerate (DD) channel \citep[e.g.,][]{iben1984}.
In the SD scenario, a white dwarf is in a binary system with 
a main-sequence star and the mass of the companion accretes to
the white dwarf. The white dwarf explodes when its mass
gets close to the Chandrasekhar mass limit.
On the other hand, the DD scenario suggests that
Type Ia SNe are caused by the merger of two white dwarfs
in a binary system. The main channel of Type Ia SNe is still unknown.

In the SD scenario, the exploding white dwarf is
surrounded by the accreting materials with a typical rate of
$\sim10^{-7}~M_\odot~\mathrm{yr^{-1}}$ \citep{nomoto1982} but
the rate is presumed to be
too small to make the recombining SNR.
Mass loss from the companion star can also make CSM around the
progenitor but the companion is likely to be a less evolved red giant
with too small mass-loss rates to establish CIE
\citep[$\sim 10^{-7}~M_\odot~\mathrm{yr^{-1}}$ or less, e.g.,][]{hachisu1999}.
However, there are rare ways to make the mass-loss rate of the system high
during the binary evolution \citep{hachisu2008} and some Type Ia SNe are
actually suggested to be a hybrid of Type Ia and Type IIn, i.e.,
Type Ia SNe exploded in a CSM as dense as those discussed in 
Section \ref{sec:iin} \citep[e.g., SN 2002ic,][]{hamuy2003}.
Thus, it is possible that
a Type Ia SN from the SD scenario evolve to a recombining SNR
but the number is expected to be very small.

In the DD scenario, we do not expect the CSM from the progenitor system
because two binary stars are white dwarfs.
However, stripped materials
at the time of the merger are suggested to remain when the merged
white dwarf explodes \citep{fryer2010}.
These materials are quite dense
($n_{ew}>10^{18}~\mathrm{cm^{-3}}$ within $r=R_\odot$,
see Figure 5 of \citet{fryer2010})
and Type Ia SNe exploded in such environment can reach
CIE and may end up with recombining SNRs. 
However, such a dense envelope 
is not obtained in a similar DD simulation of
\citet{pakmor2012}.
The fact that we do not see recombining SNRs of Type Ia SNe
may already suggest that the model obtained by \citet{fryer2010}
is not the major path to be a Type Ia SN.
Because of the uncertainty in the theoretical prediction of
Type Ia SNe from the DD scenario, we still cannot exclude the
possibility that Type Ia SNe from the DD scenario can be recombining SNRs.

To sum up, although all the recombining SNRs currently discovered are likely from
core-collapse SNe, Type Ia SNe
from both the SD and DD channels have possibility to become
recombining SNRs.
When the detailed theoretical predictions are fixed, like the existence
of the dense envelope in the DD channel, recombining SNRs
may be able
to be a probe to indicate the progenitor system of Type Ia SNe.

\section{Conclusions}
We have investigated the possible progenitors of recombining SNRs.
If a CSM which is dense enough to establish CIE in the early epochs
of the SNR evolution exists around a progenitor,
the plasma in the shocked CSM can be overionized
and the SNR can become a recombining SNR.
RSGs, especially massive ones, and Type IIn SN progenitors can
have the CSM which is dense enough
to establish CIE at the early stage of their explosions and
can evolve to recombining SNRs.
As explosions of RSGs (Type II SNe) occurs much more frequently
than Type IIn SNe, the major progenitors of recombining SNRs
are likely to be RSGs.

WR stars and white dwarfs are difficult to make recombining SNRs
with their standard mass-loss histories but they
are suggested to have mechanisms to
enhance their mass-loss rates and they can be recombining SNRs
if such mechanisms enhance their mass-loss rates.
However, these mechanisms are presumed to work on a small fraction
of these stars and such progenitors are expected to be a minor way
to have recombining SNRs.

\begin{acknowledgments}
I would like to thank the anonymous referee for the constructive comments.
The author is supported by the Japan Society for the Promotion of
 Science Research Fellowship for Young Scientists $(23\cdot5929)$.
This research is also supported by World Premier International Research
 Center Initiative, MEXT, Japan.
\end{acknowledgments}

\bibliographystyle{apj}

\end{document}